\documentclass[%
 reprint,
superscriptaddress,
nofootinbib,
 amsmath,amssymb,
]{revtex4-2}

\usepackage{amssymb}
\usepackage{amsmath}
\usepackage{amsthm}
\usepackage{mathrsfs}
\usepackage{graphicx}
\usepackage{dcolumn}
\usepackage{bm}
\usepackage[hidelinks]{hyperref}
\usepackage[usenames,dvipsnames,x11names,table]{xcolor}
\usepackage{balance}
\usepackage{dsfont}
\usepackage{tikz}
\usetikzlibrary{shapes}
\usepackage{multirow}
\usepackage{tabularx}
\usepackage{enumitem}
\usepackage{cancel,soul,ulem}
\setlist[itemize]{align=parleft,left=0pt..1em}

\newcolumntype{P}[1]{>{\centering\arraybackslash}p{#1}}

\newcommand{\Ncircle}{\raisebox{0.5pt}{\tikz{\node[draw,scale=0.7,circle,fill=orange,rotate=0](){};}}}

\newcommand{\Ecircle}{\raisebox{0.5pt}{\tikz{\node[draw,scale=0.7,circle,fill=ForestGreen,rotate=0](){};}}}

\begin{document}


\title{\textbf{Emulsions in microfluidic channels with asymmetric boundary conditions and directional surface roughness: stress and rheology}}

\author{Francesca Pelusi}
\email{f.pelusi@iac.cnr.it}
\affiliation{Istituto per le Applicazioni del Calcolo, CNR - Via dei Taurini 19, 00185 Rome, Italy}
\author{Daniele Filippi}
\affiliation{Department of Physics and Astronomy ‘G. Galilei’, University of Padova, Via F. Marzolo 8, 35131 Padova, Italy}
\author{Ladislav Derzsi}
\affiliation{Institute of Physical Chemistry Polish Academy of Sciences, Kasprzaka 44/52, Warsaw, 01-224 Poland}
\author{Matteo Pierno}
\affiliation{Department of Physics and Astronomy ‘G. Galilei’, University of Padova, Via F. Marzolo 8, 35131 Padova, Italy}
\author{Mauro Sbragaglia}
\affiliation{Department of Physics \& INFN, Tor Vergata University of Rome, Via della Ricerca Scientifica 1, 00133 Rome, Italy}


\vspace{0.5cm}
\date{\today}

\begin{abstract}

The flow of emulsions in confined microfluidic channels is affected by surface roughness. Directional roughness effects have recently been reported in channels with asymmetric boundary conditions featuring a flat wall, and a wall textured with a directional roughness, the latter promoting a change in the velocity profiles when the flow direction of emulsions is inverted [D. Filippi et al., {\it Adv. Mater. Technol.} {\bf 8}, 2201748 (2023)]. An operative protocol is needed to reconstruct the stress profile inside the channel from velocity data to shed light on the trigger of the directional response. To this aim, we performed lattice Boltzmann numerical simulations of the flow of model emulsions with a minimalist model of directional roughness in two dimensions: a confined microfluidic channel with one flat wall and the other patterned by right-angle triangular-shaped posts. Simulations are essential to develop the protocol based on mechanical arguments to reconstruct stress profiles. Hence, one can analyze data to relate directional effects in velocity profiles to different rheological responses close to the rough walls associated with opposite flow directions. We finally show the universality of this protocol by applying it to other realizations of directional roughness by considering experimental data on emulsions in a microfluidic channel featured with a flat wall and a wall textured by herringbone-shaped roughness.
\end{abstract}

\maketitle

\section{\label{sec:intro}Introduction}
\begin{figure*}[t!]
    \centering
    \includegraphics[width=1.\linewidth]{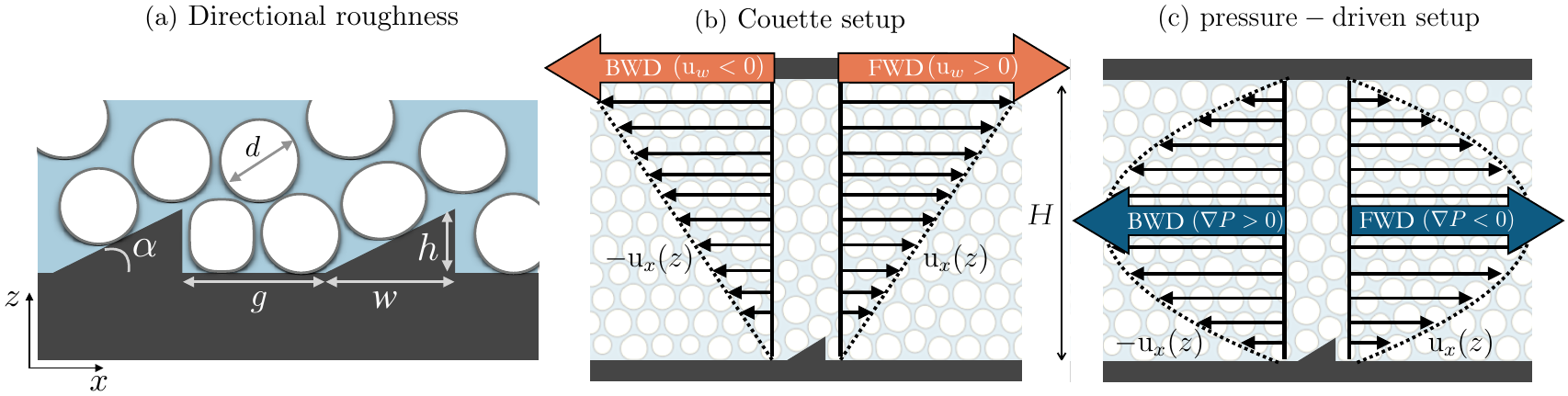} 
    \caption{Simulation setups. Panel (a): A sketch of the directional roughness realized with right-angle triangular-shaped posts. The posts are structured with a height $h$ comparable with the average droplet diameter $d$ and periodically distributed at fixed distance $\lambda = g + w$. The gap $g$ between consecutive obstacles as well as the obstacle width $w$ depend on the obstacle slope, tuned via the angle $\alpha$. All these parameters are kept fixed in simulations. Light domains refer to dispersed phase inside the continuous one. Panels (b) and (c): Sketches describing the numerical setups of a Couette (panel (b)) and a pressure-driven (panel (c)) setup. The microfluidic channel has height $H$ and asymmetric boundary conditions in both setups. Indeed, the bottom wall of the microfluidic channel is structured with a directional surface roughness (detailed in panel (a)), while the top wall is flat. The definition of forward (FWD) and backward (BWD) flows is shown in both setups.\label{fig:numericalsketch}}
\end{figure*}
Contrary to homogeneous fluids, complex structured fluids composed of soft domains, such as droplets in emulsions~\cite{Derkach09}, bubbles in foams~\cite{Ji05} or blobs in gels~\cite{Picout03}, possess a non-linear rheological response to an applied stress. This non-linear response results in a fluidization scenario that profoundly differs from that of a Newtonian fluid, comprising heterogeneous flow structures~\cite{JopMansard12,Dollet14,DolletBocher15}, shear-thinning behaviours~\cite{Pal96,Pal2000} and also the emergence of yield stress rheology for the more concentrated systems~\cite{Bonn17}. Furthermore, spatial cooperativity effects, arising from the presence of finite-sized soft domains, have been observed influencing the overall system fluidization~\cite{NicolasBarrat13,Mansard13,Scagliarini15,Dollet15}. This scenario has prompted several investigations in recent years, encompassing fundamental interests at the interface between physics, chemistry, and engineering~\cite{Mansard12,Balmforth14,Bonn17}. Beyond this fundamental importance, these materials also show a wide variety of applications, from pharmaceutics~\cite{Bouyer12} to oil recovery~\cite{Zhou19}, from cosmetics~\cite{Venkataramani20} to food  industry~\cite{Ozturk16}. The presence of interactions with the boundary conditions brings further complexities. The latter is the case -- for example -- of emulsions or foams flowing in microfluidic channels with rough walls. Indeed, surface roughness can promote droplet/bubble rearrangements close to the rough walls and alter local fluidization. These concepts were evidenced in experiments on concentrated emulsions~\cite{Goyon08,Goyon10,Mansard14} and later confirmed and further investigated in a variety of studies, both experimental and 
numerical~\cite{Derzsi17,Derzsi18,PelusiEPL19,Paredes15}. In some of these studies, microfluidic channels with asymmetric boundary conditions have been employed to address the fluidization induced by surface roughness~\cite{Derzsi17,Derzsi18}.
Roughness-induced fluidization hinges on a specific relation between the wall stress and the local shear rate close to the wall, as pointed out in Ref.~\cite{Goyon08,Goyon10}. A net difference in the slip velocity - wall stress relation close to rough and smooth surfaces was also evidenced in Ref.~\cite{Meeker04}, where flow-curves for rough and flat surfaces exhibit a difference for small stresses, with the flow-curve for flat surfaces setting below the flow-curve for rough surfaces. A similar behavior has been observed close to flat surfaces when the 
wettability of the wall changes, and the adhesion properties of the droplets/particles at the wall are altered~\cite{Paredes15,Seth12}.\\ 
The impact of the directional features of the wall texture geometry has not yet received enough attention. Recently, the role played by a {\it directional roughness} has been reported for the pressure-driven flow of concentrated emulsions within confined microfluidic channels with asymmetric boundary conditions featuring a flat wall and another wall textured by herringbone-shaped roughness~\cite{filippi2023boost}. By directional roughness, we mean a realization of the surface texture so that the geometry is different with respect to the flow direction. Considerable effects of the directional roughness on the flow properties may be coupled with the rheology of the flowing materials due to the presence of a dispersed phase. To shed light on this interplay, a detailed characterization of the rheological response of emulsions close to the rough wall is needed. In experiments, this analysis is hindered by the need for more information on the local stress field inside the microfluidic channel, which is not directly measurable in the presence of asymmetric boundary conditions since only velocity profiles are accessible~\cite{Derzsi17,Tabeling14}. This work aims to elaborate a validated protocol to reconstruct stress profiles in channels with asymmetric boundary conditions featuring a directional roughness, thus opening the possibility to relate the directional effects on velocity profiles to different rheological responses close to the rough walls when the flow direction is inverted. Specifically, we focus on a minimalist model of directional roughness in two dimensions, with the roughness structured as a series of right-angle triangular-shaped posts equally spaced (see Fig.~\ref{fig:numericalsketch}(a)) 
that make the flow direction-dependent (see Figs.~\ref{fig:numericalsketch}(b) and (c)). The simplicity of the model allows us to take full control of all parameters of the directional roughness. Numerical simulations capture different velocity profiles as well as rheological curves associated with the two flow directions. Importantly, they allow the establishment of a quantitatively validated protocol based on mechanical balance conditions to reconstruct stress profiles for pressure-driven flows in microfluidic channels with asymmetric boundary conditions. The protocol is quite universal and can also be applied to different roughness realizations. To corroborate this point, we show how the protocol can be applied to experimental data with directional roughness studied in Ref.~\cite{filippi2023boost}.\\
The paper is organized as follows: In Section~\ref{sec:methodsmaterials}, we briefly describe the numerical methodology (Section~\ref{subsec:numericalModel}) and give details on experiments (Section~\ref{sec:expSetup}); numerical results obtained in the Couette and pressure-driven setups are discussed in Section~\ref{sec:numResults}; in Section~\ref{sec:protocol}, we report on the protocol to reconstruct stress profiles from velocity data; this protocol is applied to experimental data in Section~\ref{sec:expResults}. Conclusions are finally drawn in Section~\ref{sec:conclusions}.

\section{Materials and Methods} 
\label{sec:methodsmaterials}
%

\subsection{Numerical Simulations}\label{subsec:numericalModel}
Numerical simulations were performed via lattice Boltzmann models (LBMs)~\cite{Kruger17,Succi18} using the open-source GPU-parallelized code TLBfind~\cite{PelusiCPC21}. This code specializes in simulating concentrated emulsions in microfluidic channels structured with rough walls under shear flows, so we extended it to perform pressure-driven flow simulations. TLBfind simulates two-dimensional fluids made of immiscible non-coalescing droplets via a multi-component LBM, which was observed to be optimal for simulating problems involving 
emulsions~\cite{ShanChen93,Benzi09,Dollet15,PelusiEPL19,PelusiSM21,PelusiSM23} (see Fig.~\ref{fig:numericalsketch}). The desired number of droplets and the droplet diameters are input parameters in TLBfind, making it a flexible tool for investigating the dynamics of almost monodispersed emulsions with different droplet concentrations $\Phi$. Note that the employed numerical method is a diffuse-interface method, with the interface separating the regions of the two components with a finite thickness. For this reason, we compute $\Phi$ as the fraction of domain size occupied by the dispersed phase, i.e., $\Phi = \left\{\int \int \Theta(\rho_{d}(x,z)-\rho_0)\,dx \ dz \right\}/$A, where $\rho_{d}$ is the dispersed phase density, $\rho_0$ is a reference density value, A$=H \times L$ is the domain size and $\Theta$ is the Heaviside step function (cfr. Ref.~\cite{PelusiSM21}). Notice that this protocol for the computation of droplet concentration does not ensure a one-to-one match with the experimental definition of $\Phi$. In addition, by tuning the interaction between the multi-component fluid and the walls, we impose non-adhering boundary conditions for the droplets. Further details on the method can be found in Ref.~\cite{PelusiCPC21}. \\
In this work, we simulate the flow of emulsions in a microfluidic channel of height $H/ d \sim 11$ and length $L/d \sim 40$, where $d$ is the average droplet diameter, which is kept fixed at varying droplet concentration. The microfluidic channel presents periodic boundary conditions along the $x$-direction and asymmetric boundary conditions along the $z$-direction with a top flat and a bottom rough wall (see Figs.~\ref{fig:numericalsketch}(b) and (c)). Then, the roughness on the bottom wall is structured as shown in Fig.~\ref{fig:numericalsketch}(a): the posts are right-angle triangular-shaped, having a height $h$, being equally spaced by a period $\lambda = g + w$, with a slope that can be varied via the inclination-angle $\alpha$. The latter establishes the free space length $g$ between two consecutive obstacles as well as the obstacle width $w$. The choice of $h$, $\lambda$, and $\alpha$ is the result of systematic simulations at changing the post parameters probing the optimal roughness shape to highlight the role of the directional roughness, corresponding to the case with $h/d\sim 1$, $\lambda/d\sim 10$, and $\alpha=30^{\circ}$.\\ 

For the sake of simplicity, hereafter, we name “forward flow” (FWD) the one that follows the natural slope of the roughness posts given by the angle $\alpha$. In contrast, we name “backward flow” (BWD) when the emulsion moves towards the vertical side of the posts (see Figs.~\ref{fig:numericalsketch}(b) and (c)). We perform a one-to-one comparison between a diluted and a concentrated emulsion, corresponding to a droplet concentration $\Phi = 0.384$ and $\Phi = 0.629$, respectively.  
The latter emulsion is sufficiently concentrated to show an incipient yield stress but enough fluid to flow in the microfluidic channel under a significant pressure gradient with no evidence of droplet coalescence. Furthermore, we have also been guided by the results of Ref.~\cite{filippi2023boost} showing an optimum value of $\Phi$ to observe the flow gap between FWD and BWD directions.
Velocity profiles have been estimated by applying a coarse-graining procedure at the droplet scale on averaged-in-time hydrodynamical velocity profiles.
Concerning the rheological experiment in a Couette setup, we move the upper wall with velocity ${\bm u}=u_{w}\hat{{\bm x}}$, while the lower one is immobile. The corresponding shear rate has been measured from the velocity profiles. For these simulations, we refer to the FWD flow when $u_{w} > 0$ and BWD when $u_{w} < 0$ (see Fig.~\ref{fig:numericalsketch}(b)). To explore a more complex situation with a space-dependent fluid stress profile, we simulate the flow of emulsions driven by a pressure gradient $\nabla P = \Delta P/L$ along the stream-flow direction $x$, where $\Delta P$ is the pressure difference across the channel (i.e., a pressure-driven setup). Specifically, the pressure gradient is applied on the system as a volume force with amplitude $\nabla P$. In this setup, we identify the FWD flow when the emulsion moves in a left-to-right direction (i.e., $\nabla P < 0$). In contrast, the BWD flow corresponds to the opposite situation (i.e., $\nabla P > 0$) (see Fig.~\ref{fig:numericalsketch}(c)).
Notice that the physical quantities measured with the numerical simulations are hereafter shown in lattice Boltzmann simulation units (lbu).\\
Furthermore, the TLBfind code~\cite{PelusiCPC21} allows for the local measurement of the stress tensor $\sigma_{ij}$. Specifically, for the numerical simulation data analyzed in this paper, we consider the off-diagonal component of the stress tensor $\sigma_{xz}$ which is given by the sum of a viscous contribution (i.e., second order moment of fluid LBM populations) and some elastic contributions due to interactions between the two fluid components~\cite{Dollet15,Kruger17}. In more detail, we consider the interaction between the two components, which turns into lattice forces. Thus, each lattice node interacts only with the first nearest neighbors with a specific force. These forces contribute to the local stress, which we compute as a force for the unit area (further details can be found in the Supplementary Material of Ref.~\cite{Dollet15}).
Then, after summing up all contributions, for each $z$ coordinate, the stress tensor component $\sigma_{xz}$ is averaged in time along the $x$ coordinate. This average is denoted with $\sigma(z)$. Mechanical balance implies that $\sigma(z)$ is constant for the Couette setup while it is a linear function of the $z$ coordinate in the pressure-driven setup, with a slope set by the pressure gradient~\cite{Goyon08}.

%
\subsection{Experiments}\label{sec:expSetup}
\begin{figure}[t!]
    \centering
    \includegraphics[width=.95\linewidth]
    {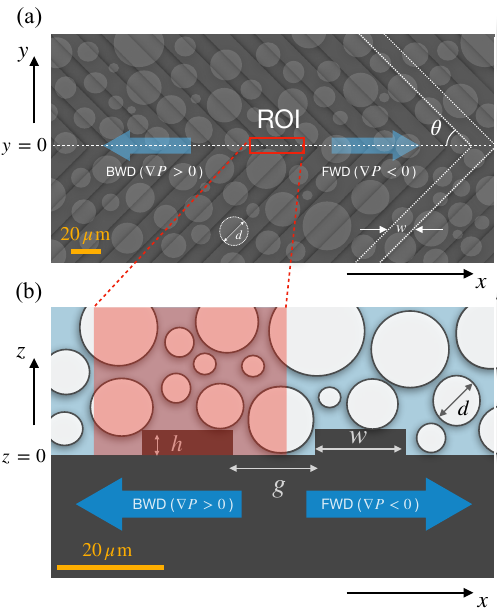}
    \caption{
    Experimental setup. Panel (a): top view sketch ($xy$-plane) of the directional roughness, textured on the bottom wall of the microfluidic channel with a herringbone design. Dotted lines highlight a representative V-shaped groove tilted at $\theta$ with respect to the flow directions. The emulsion flow is measured locally within a nearly 2D region of interest (ROI, red box) placed in the center of the channel cross section ($y=0$, horizontal dashed line) and the middle of the channel length. Panel (b): Side view ($xz$-plane) of the microfluidic channel close to the textured wall. The red rectangle is a vertical slice of the ROI. The scale bars are 2500 times shorter than the channel length $L$ and 200 times shorter than the channel width $W$, allowing for local probing of the velocity profiles. Conversely, they are only six times shorter than the channel height $H$ due to microfluidic confinement in the vertical direction.\label{fig:geomtry_exp}}
\end{figure}
\begin{figure*}[th!]
    \centering
    \includegraphics[width=.65\linewidth]{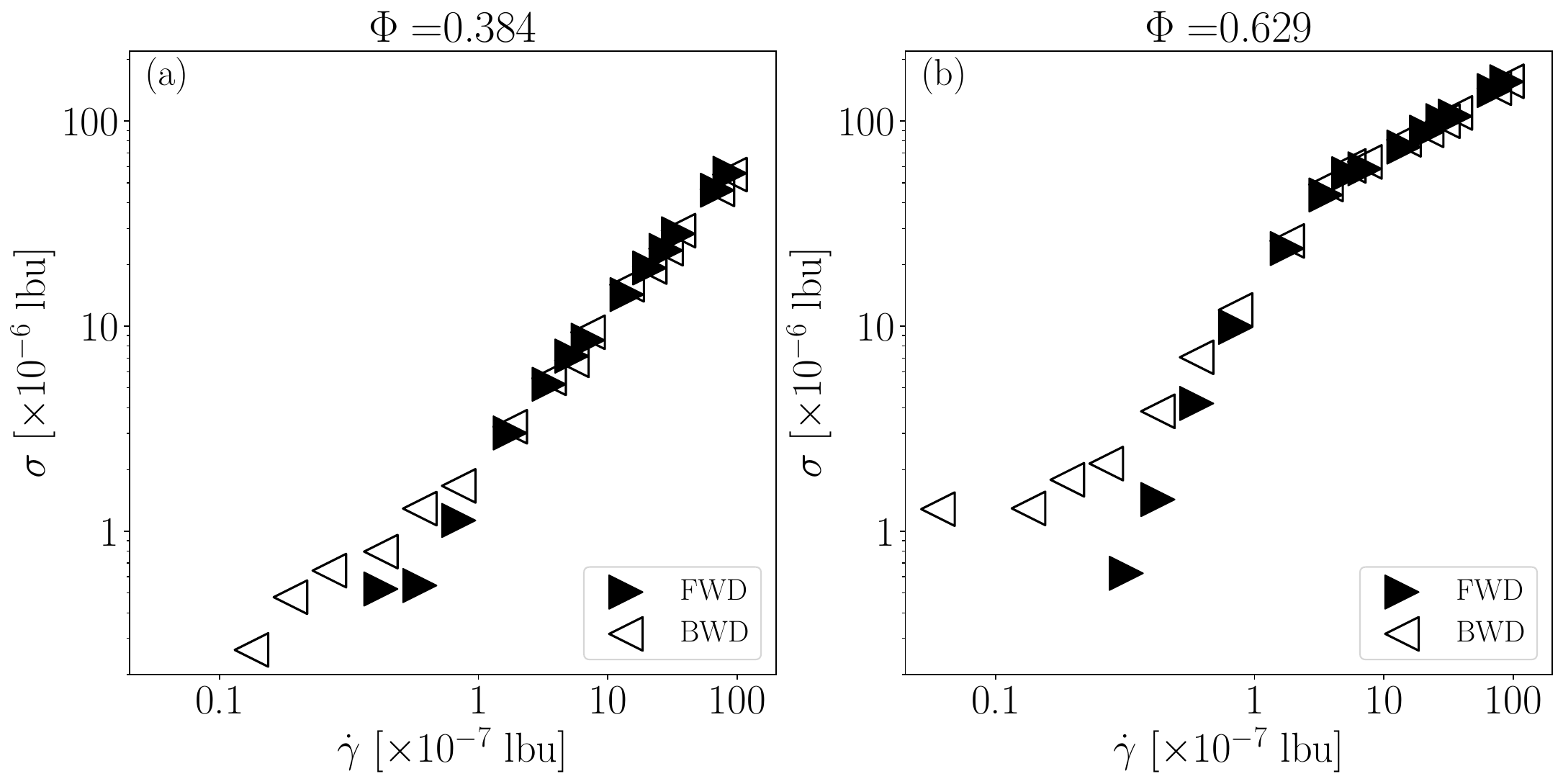} 
    \caption{Numerical results for Couette rheological measurements (see Fig.~\ref{fig:numericalsketch}(b)): shear stress $\sigma$ as a function of the shear rate $\dot{\gamma}$ for emulsions with different droplet concentration $\Phi$: a diluted ($\Phi=0.384$, panel (a)) and a concentrated emulsion ($\Phi=0.629$, panel (b)). Data for FWD ($\protect\blacktriangleright$) and BWD ({\large $\protect\triangleleft$}) flow directions are reported. \label{fig:numericalRheologyCouette}}
\end{figure*}
%
The microfluidic channel is realized by UV photolithography as described in Refs.~\cite{Derzsi17,Derzsi18,brigo08}. 
The channel is a rectangular capillary having width $W=4~\mbox{mm}$, length $L=5~\mbox{cm}$ 
and height $H=120~\mu\mbox{m}$. This provides a pressure-driven flow between extended parallel plates
of area $W \times L$
.
The bottom wall of the channel 
is textured with herringbone-shaped rectangular grooves that are textured with micrometer resolution.
%
Referring to Fig.~\ref{fig:geomtry_exp}(a), the grooves are inclined by $\theta = 45^{\circ}$ with respect to the longitudinal axis of the channel. As depicted in Fig.~\ref{fig:geomtry_exp}(b), the grooves have a rectangular shape with height $h=2.5~\mu\mathrm{m}$, width $w=17~\mu\mathrm{m}$, and they are separated from each other by a gap $g=w$. In parallel, we realize a non-textured microfluidic channel, which features both walls smooth and has the same size $L$, $W$, $H$ of the textured microfluidic channel, where the stress profile can be experimentally obtained by measuring the pressure gradient in the flow direction.
The microfluidic channel is filled with an emulsion consisting of silicone oil droplets dispersed in
a refractive index matching phase made up of a mixture of glycerin/water 54$\%$ wt/wt and stabilized by tetradecyl-trimethyl-ammonium bromide at 1$\%$ wt/wt. The procedure for emulsion preparation is described in Refs.~\cite{Goyon08,Mansard12,Derzsi17}. 
The oil droplets have an average size of $ d \simeq 9\, \mu\mbox{m}$ with a polydispersity index of 45$\%$, and occupy a volume fraction $\Phi=0.75$, at which the emulsion behaves as a yield stress fluid~\cite{seth2011micromechanical}.
The flow of the emulsion is pressure-driven by a microfluidic controller with a resolution of $\pm5$ mbar in the range of 1 to 1000 mbar (MFCS series from Fluigent, France).
The velocity profiles are measured using Particle Tracking Velocimetry (PTV) by tracking fluorescent nanoparticles whose size is $a \sim 0.2~\mu\textrm{m}$, dispersed in the glycerine/water mixture~\cite{Goyon10,Mansard14,Derzsi17,Derzsi18}.
The presence of the herringbone roughness introduces an anisotropy in the flow directions. Depending on the pressure gradient, the flows within are labeled as forward (FWD) or backward (BWD) depending on the direction towards or against the herringbone tip. As described in Ref.~\cite{filippi2023boost}, the roughness affects the entire 3D velocity profiles. However, by considering a $35\, \mu\mathrm{m} \, \times \,  7\, \mu\mathrm{m}$ region of interest (ROI) on the tip of the V-shaped grooves, thin compared to the channel width $W$, the emulsion flow can be considered to be probed over a 2D slice placed on the channel axis $y=0$ 
(see Fig.~\ref{fig:geomtry_exp}(a)). By measuring the velocity profiles of the emulsion flow within the thin, quasi-2D ROI centered around $y=0$, we provide a microfluidic device to assess the role of directional roughness and analyze the relation between stress and shear rate.

\section{Results and discussion}\label{results}

%
\subsection{Numerical simulations\label{sec:numResults}}

\begin{figure*}[ht!]
    \centering
    \includegraphics[width=.7\linewidth]{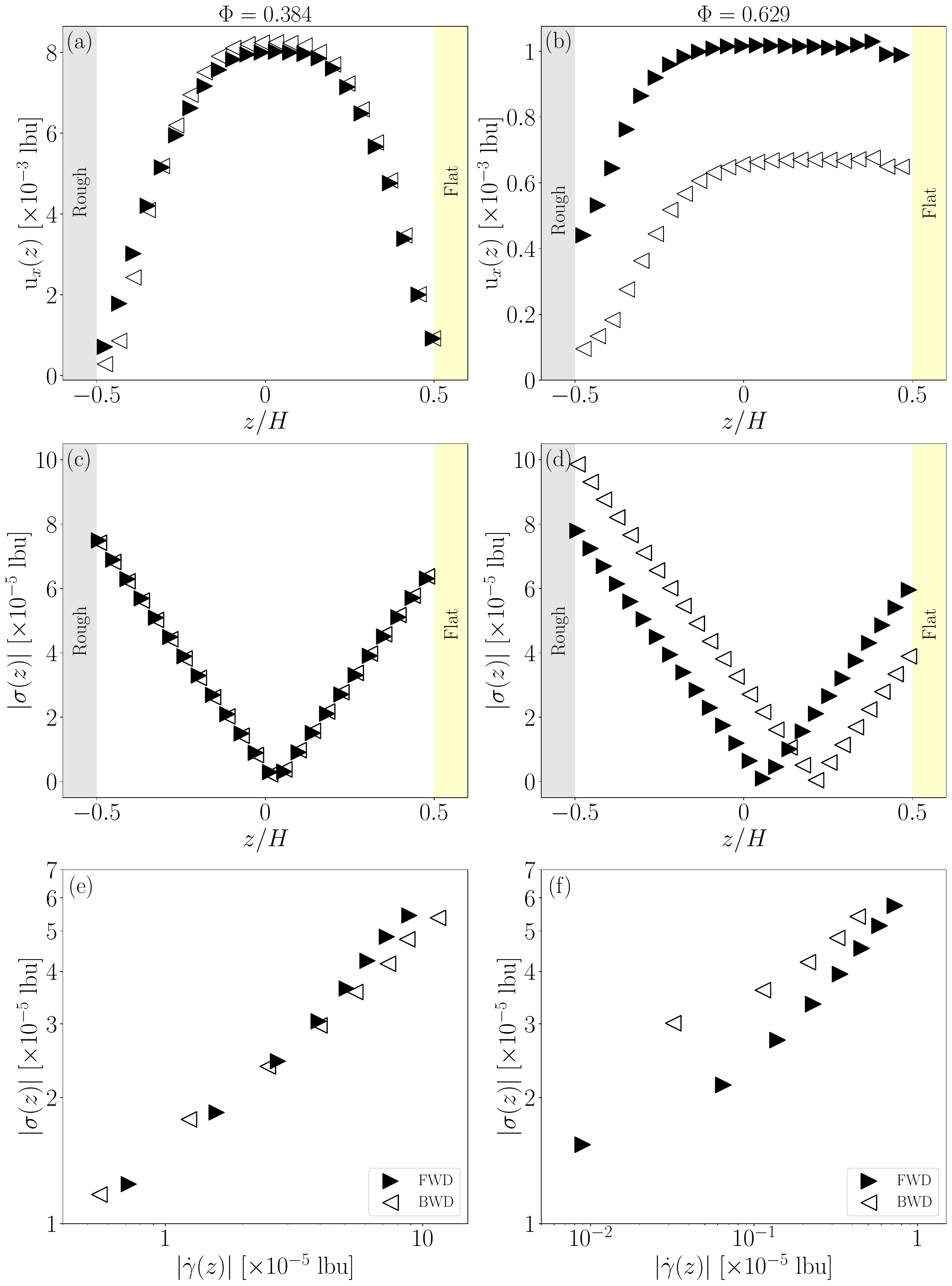}
    \caption{Numerical data for pressure-driven flows (see Fig.~\ref{fig:numericalsketch}(c)) at fixed pressure gradient $\nabla P$. In all panels, we show data for FWD ($\protect\blacktriangleright$) and BWD ({\large$\protect\triangleleft$}) flow directions. The left panels refer to the diluted emulsion with $\Phi=0.384$, while the right panels refer to the concentrated emulsion with $\Phi=0.629$. Panels (a) and (b): measured velocity profiles u$_x(z)$ as a function of the vertical direction $z$, normalized by the microfluidic channel height $H$. Panels (c) and (d): absolute value of the corresponding measured shear stress profiles $|\sigma(z)|$. Panels (e) and (f): rheological curves obtained by computing the shear rate $\dot{\gamma}$ from panels (a) and (b), respectively.
    \label{fig:numericalRheologyPoiseuille}}
\end{figure*}
%
%
First, we explored the emulsions dynamics in a Couette flow induced by the directional roughness shown in Fig.~\ref{fig:numericalsketch}(a). In this configuration, we performed a rheological characterization of a diluted ($\Phi = 0.384$) and a concentrated ($\Phi = 0.629$) emulsion, comparing the FWD and BWD directional flow (see Fig.~\ref{fig:numericalsketch}(b)). Results are reported in Fig.~\ref{fig:numericalRheologyCouette}, showing that the flow in the two directions leads to a very similar mechanical response if $\Phi$ is low enough to enhance a Newtonian behavior ($\Phi = 0.384$, Fig.~\ref{fig:numericalRheologyCouette}(a)), while the FWD and BWD flow curves differ at large values of $\Phi$ and low values of the shear rate, creating a different rheological response ($\Phi = 0.629$, Fig.~\ref{fig:numericalRheologyCouette}(b)).\\
To delve deeper into the investigation, we perform a systematic analysis in a pressure-driven setup (see Fig.~\ref{fig:numericalsketch}(c)), where the emulsions are driven by a pressure gradient $\nabla P$ and the shear stress $\sigma(z)$ across the microfluidic channel is space-dependent. Velocity profiles u$_x(z)$ as a function of the vertical direction $z$ at fixed $\nabla P$ for the same two emulsions considered in the Couette setup are shown in Figs.~\ref{fig:numericalRheologyPoiseuille}(a) and (b). In this condition, we observed that only in the case of a concentrated emulsion (Fig.~\ref{fig:numericalRheologyPoiseuille}(b)) an evident gap between FWD and BWD velocity profiles appear at fixed $\nabla P$, confirming the role of a directional roughness on the flow behavior of concentrated emulsions.
%
We now want to show that the observed gap in the flow profiles for the pressure-driven setup is rooted in a different rheological response for more concentrated emulsion triggered by the directional roughness effects. With this aim, we investigated the relation between the shear stress profile $\sigma(z)$ and the local shear rate $\dot{\gamma}(z)$. In Figs.~\ref{fig:numericalRheologyPoiseuille}(c) and (d), we show the absolute value of the measured shear stress profiles $|\sigma (z)|$ corresponding to velocity profiles in Figs.~\ref{fig:numericalRheologyPoiseuille}(a) and (b), respectively. The stress is a linear function of the $z$ coordinate, as predicted by mechanical arguments. In the Newtonian case ($\Phi = 0.384$), profiles for $|\sigma(z)|$ overlap and are symmetric with respect to the microfluidic-channel center. However, this overlap and this symmetry are lost in the more concentrated case ($\Phi = 0.629$). The latter evidence arises from the different droplet behavior close to the rough wall: in the FWD direction, droplets flow and slide on the obstacle slope and ``jump" beyond the post, thus also triggering a vigorous plastic activity; contrariwise, in the BWD direction, droplets hit a vertical obstacle and remain blocked, undergoing a slowdown and a local deformation which turns into a more significant stress value. A more quantitative analysis is discussed in the Supplementary Material. Furthermore, notice that, despite an evident shift in the stress profiles between FWD and BWD direction, the droplet concentration remains almost uniform across the channel section, with a slightly larger value close to the rough wall in the BWD direction, which is symptomatic of the droplet blockage when entering in contact with the vertical posts (see Supplementary Material). To make progress, we needed to estimate the local shear rate $\dot{\gamma}(z)$. In the most concentrated case, velocity profiles present some noise due to the simulation resolution, thus making a precise measurement of $\dot{\gamma}(z)$ difficult. For this reason, a filtering procedure performed with a polynomial function is needed to compute the velocity gradient and extrapolate the corresponding rheological data. The resulting relation between $|\sigma (z)|$ and the absolute value of the local shear rate $|\dot{\gamma}(z)|$ is shown in Figs.~\ref{fig:numericalRheologyPoiseuille}(e) and (f) for the diluted and concentrated emulsion, respectively, confirming the evidence of a different surface rheological response induced by a directional roughness also in a pressure-driven setup. Notice that, since the asymmetric boundary conditions lead to asymmetric velocity and stress profiles with respect to the center of the microfluidic channel and we are interested in investigating the rheological properties of emulsions close to the rough wall, we consider only the values measured in the microfluidic channel region close to the rough wall where $\sigma (z)>0$. Further, we do not consider the contribution of one droplet boundary layer close to $z = -H/2$ because data for BWD flows are affected by a concavity change in that region.\\ 
The analysis of data shown in Figs.~\ref{fig:numericalRheologyPoiseuille}(e) and (f) is possible because we can directly measure the local stress in the simulations. We now want to devise a quantitatively validated protocol for the reconstruction of the stress profile indirectly from the velocity profiles, i.e., in those situations where the stress is not locally measurable but the velocity profile is. The latter may be the case of real experiments of emulsions driven by a pressure gradient in microfluidic channels with directional roughness in the presence of asymmetric boundary conditions at the channel walls~\cite{filippi2023boost}. With some validated protocol to reconstruct the stress profiles from experimental velocity profiles, we can then consider data from experiments and see if the picture portrayed by  Figs.~\ref{fig:numericalRheologyPoiseuille}(d) and (f) still holds.
\subsection{Stress profile in channels with asymmetric boundary conditions: a validated protocol from numerical simulations}\label{sec:protocol}
%
%
\begin{figure}[t!]
    \centering
    \includegraphics[width=.9\linewidth]
    {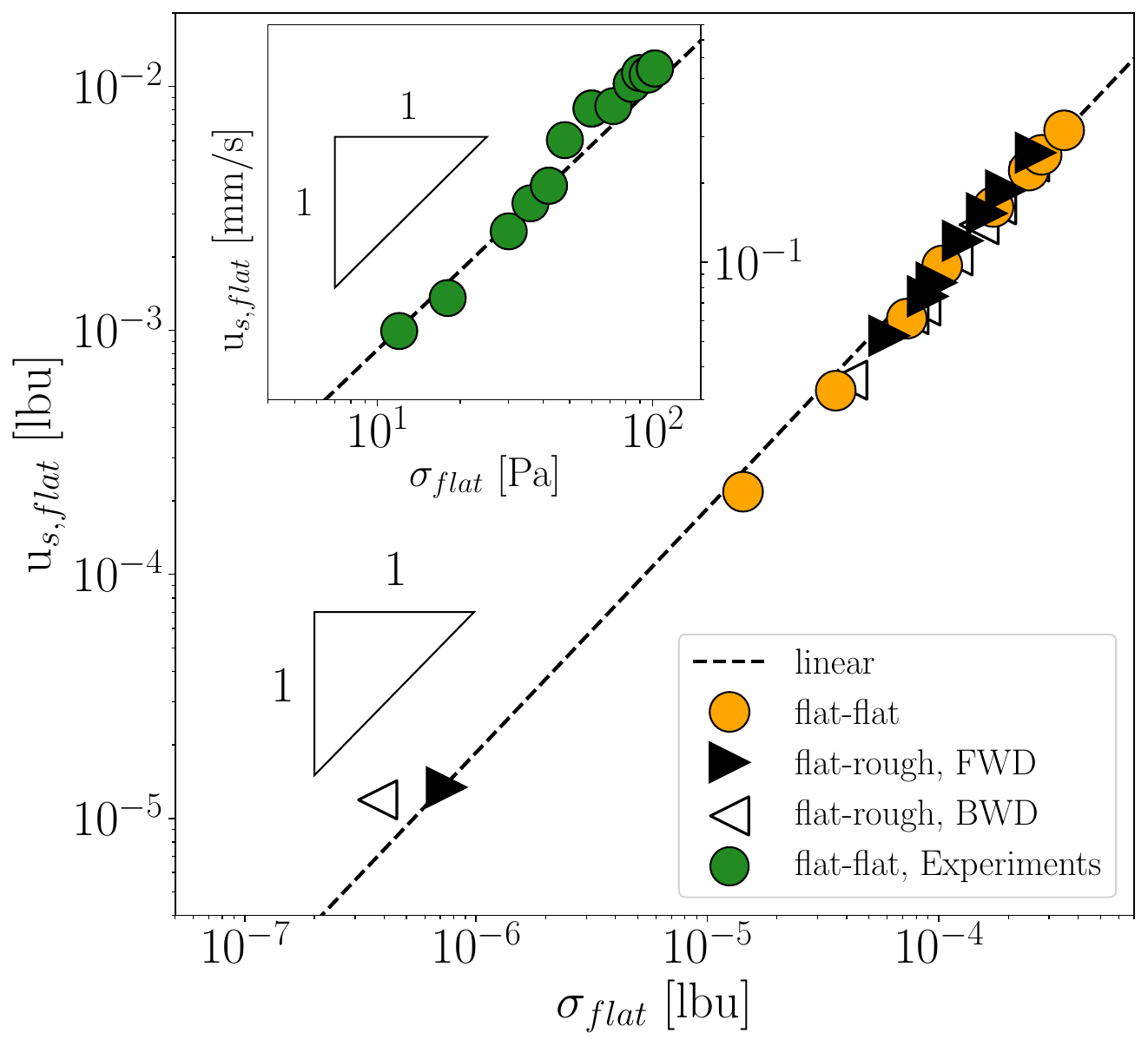}
    \caption{Slip velocity u$_{s, flat}$ as a function of the shear stress $\sigma_{flat}$ measured in the proximity of the flat wall. Data for FWD ($\protect\blacktriangleright$) and BWD ({\large$\protect\triangleleft$}) flow directions are compared with data for the case of a symmetric flat-flat microfluidic channel ($\protect\Ncircle$). All data collapse on a linear scaling (dashed lines), in agreement with the non-adhering case in Ref.~\cite{Seth12}. In the inset,  experimental slip velocity as a function of the corresponding wall shear stress is reported ($\protect\Ecircle$) for the experimental emulsion at $\Phi=0.75$ driven in a symmetric flat-flat microfluidic channel having the same size $W= 4 \, \mbox{mm}, L=5 \,\mbox{cm}, H=120 \, \mu\mbox{m}$, at different pressure gradients from $\nabla P = L^{-1}100$ mbar to $\nabla P = L^{-1}850$ mbar (see Section~\ref{sec:expResults}).\label{fig:vslip_stress_relation}}
\end{figure}
\begin{figure*}[t!]
    \centering
    \includegraphics[width=1.\linewidth]{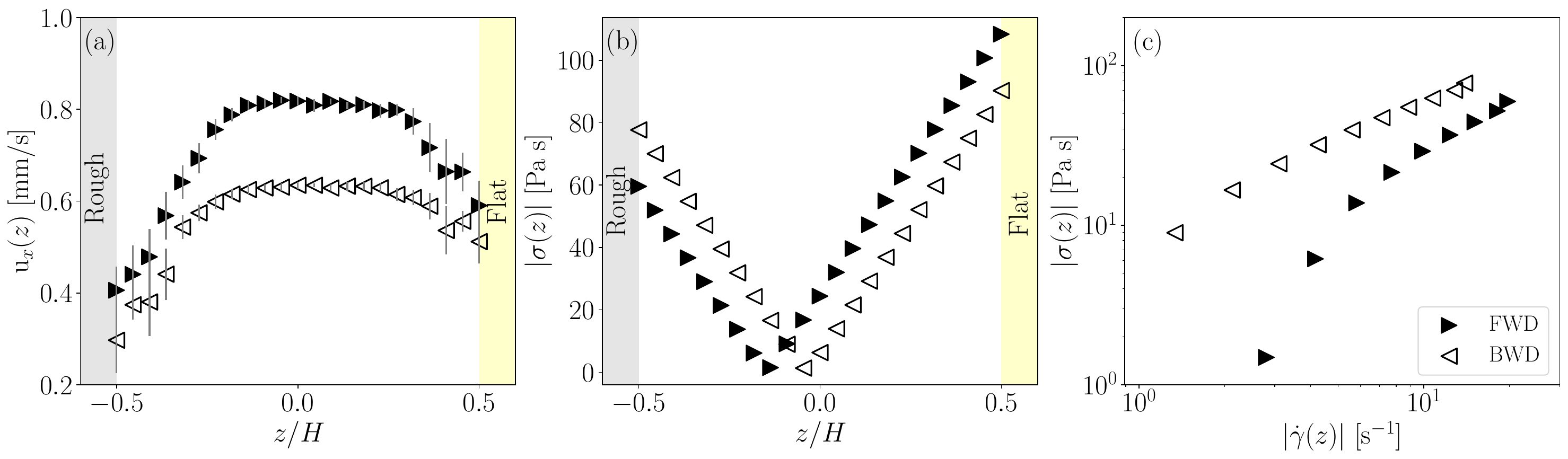}
    \caption{
    Experimental data. FWD ($\protect\blacktriangleright$) and BWD ({\large$\protect\triangleleft$}) flow directions for the emulsion at $\Phi=0.75$ in the flat-rough microfluidic channel, driven by $\nabla P=L^{-1}\,700$\,mbar. Panel (a): Experimental velocity profiles u$_x(z)$ measured by PTV as a function of the vertical direction $z$, normalized by the height $H$ of the microfluidic channel. Panel (b): Absolute value of the stress profile $|\sigma(z)|$ within the microfluidic channel obtained from Eq.~\ref{eq:protocol} by running the wall slip-stress protocol described in Section~\ref{sec:protocol}. Panel (c): Flow curves extracted from the stress data reported in (b) as a function of shear rate $\dot{\gamma}$ extracted from panel (a). \label{fig:experiments}}
\end{figure*}
%
The starting point is the mechanical-balance condition $\partial_z \sigma(z)= \nabla P$, implying that the stress profile across the microfluidic channel is linear, with a slope that is given by the pressure gradient $\nabla P$. The missing information to reconstruct the full profile $\sigma(z)$ is the stress value in some given location: we choose the stress value on the flat wall $\sigma_{flat}$ because if we know the relation between the slip velocity at the flat wall (u$_{s,flat}$) and the corresponding wall-stress $\sigma_{flat}$, the problem is closed. To estimate the relation between u$_{s,flat}$ and $\sigma_{flat}$ one can consider a pressure-driven flow in a flat-flat microfluidic channel (no roughness); in this case, the analytical solution for the stress profile predicts a linear profile that is symmetrical with respect to the microfluidic-channel center: the value of $\sigma(z)$ on one of the two flat walls is $\sigma(z=\pm H/2) = \sigma_{flat} = \mp (\nabla P) H/2$. Thus, by collecting measurements of u$_{s,flat}$ for different values of $\nabla P$, we can obtain a relation between u$_{s,flat}$ and $\sigma_{flat}$. We show the corresponding data for the concentrated emulsion ($\Phi = 0.629$) in a flat-flat channel as orange circles in Fig.~\ref{fig:vslip_stress_relation}. These data are in agreement with the experiments of Seth and coworkers~\cite{Seth12} for concentrated emulsions in a Couette rheometer with flat and non-adhering walls, showing that the slip velocity u$_{s,flat}$ scales linearly with the wall-stress $\sigma_{flat}$. Notice that no {\it a priori} prediction can be stated for estimating the constitutive laws relating wall stress and slip velocity: despite the linear behavior that would be expected for such boundary conditions, we cannot predict at this stage the scaling prefactor. If the stress-velocity relation $\sigma_{flat}($u$_{s,flat})$ is specific to the coupling between the emulsion and the wall, then we expect to find the very same relation if we consider the emulsions in flat-rough microfluidic channels and we look at the stress-velocity relation close to the flat wall. This is a non-trivial point in that it validates the protocol that allows us to assign the stress at the flat wall in a flat-rough microfluidic channel once we know the relation between stress and slip velocity in a flat-flat channel. Numerical data reported in Fig.~\ref{fig:vslip_stress_relation} confirm the expectations for both FWD and BWD flow directions. \\
In summary, to measure the stress profile from velocity data in a pressure-driven microfluidic channel with asymmetric (i.e., flat-rough) boundary conditions, we need to follow these steps: {\it i)} we can first estimate the relationship between the slip velocity and the stress on the flat wall $\sigma_{flat}($u$_{s,flat})$ from dedicated experiments in microfluidic channels with symmetric boundary conditions (both flat walls); {\it ii)} the mechanical-balance condition implies a linear stress profile $\sigma(z)$ as a function of the vertical position $z$ with a slope given by the pressure gradient $\nabla P$; {\it iii)} from the measure of the slip velocity at the flat wall (u$_{s,flat}$) in experiments in microfluidic channels with asymmetric boundary conditions, we compute $\sigma_{flat}($u$_{s,flat})$ from the relation obtained in {\it i)}, thus determining the stress profile as:
\begin{equation}
    \sigma(z) = - \sigma_{flat} - \nabla P \left( z-\frac{H}{2}\right).
    \label{eq:protocol}
\end{equation} 
In the next section, we apply the operative protocol outlined above by points {\it i)-iii)} to experimental data collected in a microfluidic channel textured by a directional roughness consisting of herringbone-shaped grooves~\cite{filippi2023boost}.
\subsection{Protocol application to experiments with a different directional roughness}\label{sec:expResults}
To address the universality of the protocol described in Section~\ref{sec:protocol} to reconstruct the stress profiles in microfluidic channels with different directional roughness realizations, we considered a microfluidic channel with flat-flat walls separated by the same height $H$ of the flat-rough microfluidic channel shown in Fig.~\ref{fig:geomtry_exp}.
Notice that the geometry of the directional roughness addressed by the experiment differs from that of the numerical simulation. Nevertheless, the protocol that we have numerically validated is based on the mechanical balance hence it is expected to hold also on different roughness realizations.\\
In the inset of Fig.~\ref{fig:vslip_stress_relation}, we report the slip velocity measured in a flat-flat channel (u$_{s, flat}$) as a function of wall stress $\sigma_{flat}= L^{-1} \Delta P \,(\pm H/2)$. Data refer to an emulsion with droplet concentration $\Phi=0.75$ for different pressure gradients. The values of slip velocities are the average over values measured on both flat walls to avoid noise fluctuations. The resulting wall slip velocity-stress relation is as linear as obtained from numerical data for a flat-flat channel. The latter data show that the wall slip is proportional to the wall stress on the flat wall. We then apply the protocol described in Section~\ref{sec:protocol} to our experimental data: we first measured u$_{s,flat}$ from velocity profiles reported in Fig.~\ref{fig:experiments}(a) in the flat-rough microfluidic channel (described in Section~\ref{sec:expSetup}) at
fixed pressure gradient
$\nabla P = L^{-1}\,700$ mbar; then, we estimated the corresponding value of $\sigma_{flat}$ from the inset of Fig.~\ref{fig:vslip_stress_relation}; finally, we applied the key protocol steps discussed in Section~\ref{sec:protocol} to obtain the stress profiles $\sigma(z)$. Fig.~\ref{fig:experiments}(b) reports the resulting absolute value of the shear stress profiles $|\sigma(z)|$ for FWD and BWD flows shown in Fig.~\ref{fig:experiments}(a), and it confirms that the existence of different slip velocities on the flat wall results in a shift between the FWD and the BWD shear stress profiles. 
The experimental flow curves shown in Fig.~\ref{fig:experiments}(c) are achieved by plotting $|\sigma (z)|$ as a function of the absolute value of the shear rate $|\dot{\gamma}(z)|$ obtained as the derivative of the velocity profile $\dot{\gamma}(z)=\partial_z $u$(z)$. Also in experiments, we focus only on the half of the channel close to the rough wall, i.e., the spatial region where $\sigma(z) > 0$, since we are interested in investigating the rheological properties close to the rough wall. Flow curves in Fig.~\ref{fig:experiments}(c) verify the emergence of different rheological responses in the pressure-driven flow with a different directional texturing. This result further suggests that the directional roughness geometry triggers a different rheological response between FWD and BWD flow directions due to the different 
directional stress on the droplets. Notice that it has been essential to perform a fitting procedure on velocity profiles to obtain experimental flow curves; otherwise, the noise introduced by the numerical derivatives of the raw velocity data is so high that it masks the stress difference of the FWD and BWD flow curves.
%

\section{Conclusions}\label{sec:conclusions}
%
Emulsions flowing in microfluidic channels with rough walls can exhibit directional roughness effects~\cite{filippi2023boost}. This scenario manifests when the realization of the wall roughness produces different flow profiles depending on the flow direction. When the textures are arranged asymmetrically, a protocol is needed to reconstruct stress profiles inside the microfluidic channel for pressure-driven flows. This protocol is key to studying how the emulsion rheology close to the rough walls depends on the directionality of the flow. In this paper, we established this protocol with a minimalist model, featuring lattice Boltzmann numerical simulations of emulsions moving within a 2D microfluidic channel textured at one wall with right-angle triangular shaped posts, using the open-access TLBfind code~\cite{PelusiCPC21}.
We show that directional effects on the velocity field are associated with the onset of a different rheological response in the two flow directions at large droplet concentrations. Numerical simulations are essential to develop an operative protocol to obtain the stress profile inside the microfluidic channel starting from the velocity profile of pressure-driven flows. The robustness of this protocol hinges on the underlying mechanical balance condition. To show the universal application of the presented protocol to other directional roughness realizations, we applied the protocol to experimental data taken on a microfluidic channel textured by herringbone roughness, considering the emulsion flow at a volume fraction above the yielding on the tip of the `V'-shaped grooves~\cite{filippi2023boost}.\\ 
Notably, as reported by Fig.\ref{fig:numericalRheologyCouette}(b), Fig.\ref{fig:numericalRheologyPoiseuille}(f), Fig.~\ref{fig:experiments}(c), at a fixed shear rate, the shear stress is more significant in the backward direction than in the forward direction. This rheological difference echoes previous findings reporting larger shear stress values in a concentrated emulsion when flowing through a rough microfluidic channel compared to a flat one~\cite{Goyon10}. It also echoes the observations of a different rheological response when boundary hydrophobicity is changed and an increase in shear stress is observed with a hydrophobic wall compared to a hydrophilic one~\cite{Paredes15,Seth12}. However, in the present study, the rheological response is not due to a change of the wettability pattern on the wall but rather by a change of flow direction with respect to a fixed suitable physical texture of the wall roughness. Notice that the developed protocol for stress estimation proves an invaluable tool for experimentally estimating the stress within a microchannel with asymmetric boundary conditions on opposite walls.\\
Resuming, we confirm that boundary conditions consisting of textures asymmetrically arranged with respect to the flow direction can induce different rheological responses resulting in different macroscopic flow properties. As expected, either the bulk rheology or the presence of a dispersed phase with a characteristic size comparable to the roughness plays a crucial role in setting the local stress close to the asymmetric rough elements. This property can be fruitfully exploited to design a convenient geometry to induce a local change in the rheological response of the flows.\\ 
Data on directional roughness are scarce in the literature; therefore, there are many perspectives for future advancements. For example, it would be interesting to understand the optimal design of a directional roughness that allows for passive control of the response of the emulsions in microfluidic channels. This obviously depends on the micro-mechanics of the emulsion droplets close to the wall, and in this perspective a synergistic work between experiments and numerical simulations could help to shed light on the matter. Furthermore, a detailed characterization of the role played by the pressure gradient in the pressure-driven setup may better clarify the parameter combination that maximizes the directional roughness effect on emulsion flow. Finally, it would be interesting also to investigate the stress-slip relationships as a function of the shape of the roughness in a corresponding symmetric scenario, i.e., with a microfluidic channel patterned with two identical directional roughness. In this case, the general protocol explained in Sec.~\ref{sec:protocol} is trivial because the channel symmetry establishes the stress profile to cross zero in the center of the channel, but the velocity profiles and slip velocities may differ because the emulsion interacts with a different roughness realization in the two directions.

\section*{Acknowledgements}
This work has been carried out within the TEXTAROSSA project (G.A. H2020-JTI-EuroHPC-2019-1 No. 956831). Partial funding from BIRD 2021 “BiodivSeq” of Padua University is kindly acknowledged.  D.F. acknowledges funding from Programma Operativo Nazionale (PON) “Ricerca e Innovazione” 2014–2020 (ReactEU-RUAPON-Viganov) of the Italian Ministry of University and Research.  MS gratefully acknowledges the support of the National Center for HPC, Big Data and Quantum Computing, Project CN\_00000013 - CUP E83C22003230001, Mission 4 Component 2 Investment 1.4, funded by the European Union - NextGenerationEU.

\bibliographystyle{rsc}
\bibliography{export}

\end{document}